# Exploring Nanoscale Ferroelectricity in Doped Hafnium Oxide by Interferometric Piezoresponse Force Microscopy

Liam Collins[†], and Umberto Celano[ζ]

[†]Center for Nanophase Materials Sciences, Oak Ridge National Laboratory, Oak Ridge, Tennessee 37831, USA

[ζ]IMEC, Kapeldreef 75, B-3001 Heverlee (Leuven), Belgium

Keywords: Hafnium oxide, Ferroelectricity, Piezoresponse force microscopy, Interferometric displacement sensing.

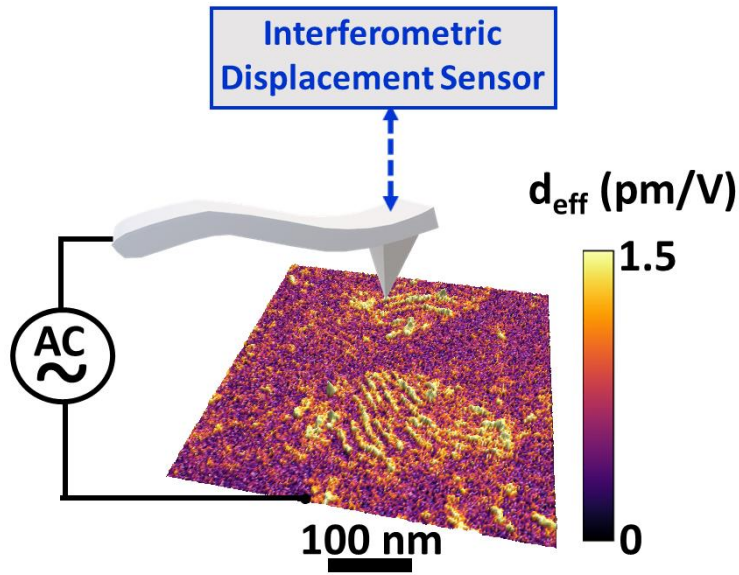

ABSTRACT

Hafnium oxide (HfO$_2$)-based ferroelectrics offer remarkable promise for memory and logic devices in view of their compatibility with traditional silicon CMOS technology, high switchable polarization, good endurance and thickness scalability. These factors have led to steep rise in research on this class of materials over the past number of years. At the same time, only a few reports on the direct sensing of nanoscale ferroelectric properties exist, with many questions remaining regarding the emergence of ferroelectricity in these materials. While piezoresponse force microscopy (PFM) is ideally suited to probe piezo- and ferro-electricity on the nanoscale, it is known to suffer artifacts which complicate quantitative interpretation of results and can even lead to claims of ferroelectricity in materials which are not ferroelectric. In this paper we explore the possibility of using an improved PFM method based on interferometric displacement sensing (IDS) to study nanoscale ferroelectricity in bare Si doped HfO$_2$. Our results indicate a clear difference in the local remnant state of various HfO$_2$ crystallites with reported values for the piezoelectric coupling in range 0.6-1.5 pm/V. In addition, we report unusual ferroelectric polarization switching including possible contributions from electrostriction and Vegard effect, which may indicate oxygen vacancies or interfacial effects influence the emergence of nanoscale ferroelectricity in HfO$_2$.

INTRODUCTION

Hafnium oxide (HfO$_2$) based ferroelectrics are one of the most studied binary oxide systems for applications in non-volatile memory applications.[1] Interest arises from promising compatibility of these materials with the existing complementary metal oxide semiconductor (CMOS) process technology. Moreover, when compared with the traditional alternatives for integrated ferroelectric devices,[2] ferroelectric HfO$_2$ (FE-HfO$_2$) is lead-free and offers improved thickness scaling at reduced leakage with low power switching, with potential applications in memory[3-5] and logic.[6] Despite the remarkable improvements in device processing and performance over the past number of years, many questions regarding the origins of ferroelectricity remain elusive. Most experimental testing involve ferroelectric capacitors and require the presence of electrodes. Relevant approaches include temperature-dependent dielectric permittivity measurements, second harmonic generation, as well as classical polarization–electric field ($P$–$E$) hysteresis measurements with macroscopic electrodes. While macroscopic characterization techniques on electrode devices help our understanding regarding global ferroelectric behavior, however, these measurements provide little knowledge on possible nanoscale origins of ferroelectricity in FE-HfO$_2$. Unfortunately, such macroscopic experiments are not entirely suitable for exploring polycrystalline FE-HfO$_2$ films which are expected to have heterogenous ferroelectric behavior. This necessitates tools capable of exploring intrinsic ferroelectric on the length scales of the local inhomogeneities themselves (e.g. phases, domains, defects). Therefore, a full understanding of the switching processes in these devices, requires the development of reliable experimental techniques for probing ferroelectricity at the relevant scale.

Piezoresponse force microscopy (PFM)[7] is the gold standard for exploring piezo- ferro-electric behavior on the nanoscale. Imaging and manipulating of ferroelectric domains and probing polarization dynamics has been achieved with the emergence of PFM, offering the advantages of probing nanometer scale volumes and high-resolution imaging and spectroscopy.[8] PFM is a scanning probe microscopy (SPM)-based technique which measures the dynamic electromechanical response of the ferroelectric sample when an ac voltage is applied to the SPM tip in mechanical contact with a surface.[9] However, to date only limited success has been achieved and conclusively evidence of ferroelectricity on the bare surface of FE-$HfO_2$ using PFM is distinctly missing.

Regards investigation of fundamental understanding of FE-$HfO_2$, significant efforts have been reported including ab initio simulations[10] and direct observation with transmission electron microscopy (TEM).[11] Progress relating to PFM on Hafnia has seen the work of Stolichnov et al., using non-resonance PFM on the surface of capacitors for the analysis of individual polarization domains through the top electrode that provides a uniform electric field application in the $Hf_xZr_{(1-x)}O_2$.[12] On the same material system, Chouprik et al., correlated the information of domains sensing by PFM with structural analysis by TEM, observing the domain structural evolution during switching.[13] Buragohain et al. focused on the kinetics of domain nucleation and wall motion, reporting on the use of PFM in conjunction with pulse switching measurements for the nanoscopic visualization of domains evolution during wake up in La-doped $HfO_2$.[14] Recently, using a combination of PFM and second harmonic generation, Cheema et al. demonstrated ferroelectricity in ultrathin (1 nm thick) $Hf_{0.8}Zr_{0.2}O_2$ (HZO) when using a confinement layer (TiN), whilst also highlight the importance of quantitative PFM measurements.[15]

This work, therefore, focuses on the overcoming barriers to extracting local information on ferroelectricity in $HfO_2$ using a recently developed interferometric PFM. We demonstrate the interferometric PFM can be used to quantify local electromechanical coupling and ferroelectric behavior in $HfO_2$. Using both nanoscale imaging and spectroscopy we report an anonymously small effective piezoelectric coupling coefficient ($d_{eff}$ of < 2 pm/V) in Si-doped $HfO_2$. We demonstrate the appearance of nanoscale domains and ferroelectric switching, as well as additional contributions to due to possible electrostrictive or Vegard effects (e.g. oxygen vacancy motion or interface effects). Our results impact the community twofold, first they provide information to engineers and material scientists actively trying to control and harness ferroelectricity in binary oxides, and second, they demonstrate the applicability of interferometric PFM for the investigation of ferroelectric effects in materials with ultra-low piezoelectric coefficient.

RESULTS AND DISCUSSION

We focus this study on silicon doped $HfO_2$ ($Si:HfO_2$), that is a widespread material system with demonstrated uses as active layer in multiple ferroelectric devices.[16, 17] Standalone capacitors are fabricated on a $p^+$-Si substrate ($10^{19}$ at/cm$^3$), where 8 nm of FE-$HfO_2$ are deposited by atomic layer deposition (ALD) at 300 ºC. Silicon is used as dopant during the oxide deposition, we select a concentration in range of 3.6 mol. %. The stoichiometry is confirmed by Rutherford back scattering (RBS) with elastic recoil detection (ERD), not shown. After the oxide deposition, 50 nm thick Si capping layer is deposited by physical vapor deposition and the stack is annealed using a 1000 ºC for 30 seconds. This step induces the oxide crystallization leading to a polycrystalline film containing a mixture of crystallites with different phases such as monoclinic, tetragonal, cubic and orthorhombic (m-, t-, c-, and, o-phase). Afterward, the Si capping layer is patterned with reactive ion etching thereby creating individual capacitors for electrical measurements and in between them, exposed regions where PFM can be performed. Cross-sectional HRTEM image of the capacitor is shown in the inset of Figure 1a, confirming the layer thickness, the sharp interfaces and the polycrystalline nature of the film. Conventional P-V hysteresis and I-V loops are acquired on capacitors with size 100 x 100 µm² and shown in Figure 1b. Similarly, Figure 1c shows the leakage current sensed on the oxide surface using conductive atomic force microscopy (C-AFM). To summarize, electrically extracted values for remnant polarization ($P_r$) and coercive field ($E_c$) are 20 µC/cm² and 2.6 MV/cm respectively. While the C-AFM results indicate a clear rectifying I-Vs behavior induced by the asymmetric stack, with a low electrical leakage in the voltage range +/- 4 V.

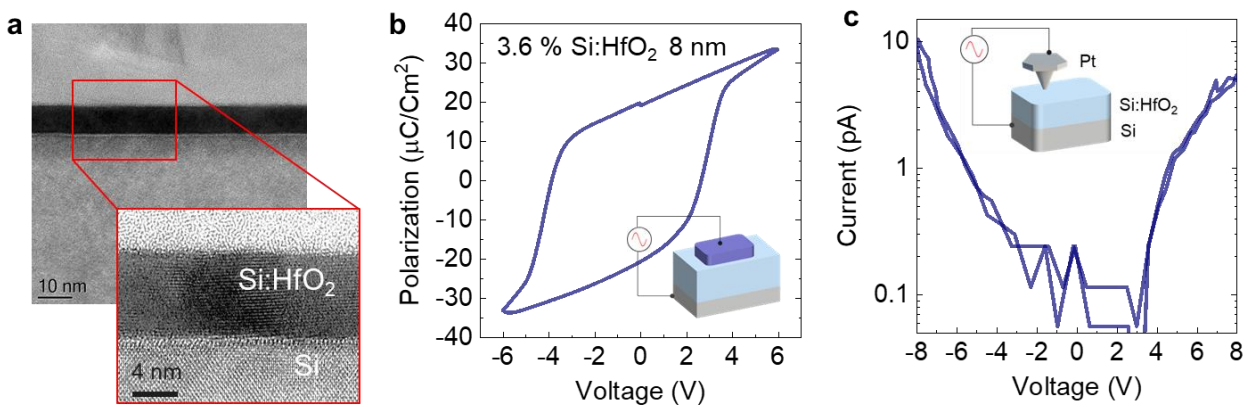

**Figure 1,** (a) HRTEM cross-section of the FE capacitor, 8 nm FE-HfO₂ is sandwiched between the Si and Poly-Si top electrode. (b) Polarization-voltage (P-V) of 100 x 100 µm² capacitor and (c) I-V spectroscopy performed using C-AFM on the FE-HfO₂ surface.

To explore the emergence of nanoscale ferroelectricity we interrogate the bare film using PFM. In PFM an ac voltage *V* applied to a probe tip held in contact with the sample resulting in a periodic surface displacement. The surface displacement arises due to the converse piezoelectric effect and is most commonly measured using the optical beam deflection (OBD) or "beam bounce" method.[18, 19] OBD is an indirect measurement such that the measured signal is proportional to angular changes of the cantilever, rather than the displacement of the tip. Thus, for quantitative PFM additional calibration of the deflection sensitivity (i.e. m/V) is needed to relate the measured angular deflection signal to the surface displacement of interest. Unfortunately, this requires assumptions about the cantilever mode shape which often fail, especially under dynamic contact mode measurements.[20]

In an ideal "vertical" PFM measurement, the out of plane piezoelectric tensor component (i.e. $d_{33}$) normal to the surface can be quantified directly. In reality the true surface displacement can be influenced by additional tensor components (e.g. in plane components along the surface normal) and or clamping effects, therefore it is more common to discuss PFM measurements in terms of the effective piezoelectric coefficient ($d_{eff}$).[2, 21] Since PFM is a dynamic measurement lock in amplifiers are typically employed to detect the magnitude of $d_{eff}$, while the phase of the response provides information about the polarization direction. The $d_{eff}$ is commonly reported in values of pm/V by dividing the measured tip displacement by the amplitude of the tip-sample voltage.

From topography channel, Figure 2a, the sample oxide surface was found to be relatively smooth having a roughness value of ~200 pm. We do observe regions across the sample surface which appeared slightly raised, which is not surprising considering this is a polycrystalline film (see Figure S2). As a first attempt we employed single frequency PFM measurements on the bare HfO$_2$ surface. This method was found to be unsuitable for quantifying $d_{eff}$ (Figure S1). The measurement lacked sufficient signal to noise performance at low frequencies (i.e. below the contact resonance frequency). At higher frequencies the contrast was dependent on proximity to the resonance peak (i.e. dependent on cantilever mode shape and not surface displacement). This is due to the fact that single frequency PFM methods are particularly sensitive to artifacts (e.g. indirect topographical crosstalk[22]) when operated at frequencies in the vicinity of the contact resonance peak. Meanwhile, the necessity for resonant amplification on the bare $HfO_2$ would appear to make single frequency PFM wholly unsuitable due to an apparent low $d_{eff}$.

To avoid such effects and gain sensitivity we first explore the use of band excitation (BE)-PFM.[23,24] In contrast to traditional single frequency PFM, BE-PFM is a multifrequency approach allowing improved signal to noise through operation close to cantilever vibrational modes.[23] At the same time, BE-PFM allows for a complete characterization of the contact mechanics by tracking the behavior of contact resonance peak at each spatial location. In Figure 2 we examine the $d_{eff}$ and contact resonance frequency shift images collected on the bare oxide. These images were determined from fitting coefficients of a similar harmonic oscillator (SHO) fit of the resonant peak at each spatial location. In supplementary Figure S3 we provide images for all the SHO fit parameters as well as the $R^2$ goodness of fit. Operating on the bare surface, as opposed to top electrodes, we see interesting contrast in the piezoelectric properties. We observe a finite response across the entire sample surface, having an average $d_{eff}$ of 3.8 ± 0.7 pm/V, consisting of localized domains (~100s nm) of higher (~4 pm/V) and lower (~2.7 pm/V) piezoelectric coupling formed on length scales of 100s nm. In addition, from the high-resolution piezoresponse images (Figure 2b) we observed periodic structures on the length scales of 10s of nm (highlighted with red arrows). Such contrast is of significant relevance to exploring emergence of ferroelectricity on the nanoscale in this material system. From Figure 2c, we also observed localized variation in the contact resonance frequency channel. Contrast in the frequency channel indicates conservative changes in tip-sample contact mechanics. The frequency shift, $d_{eff}$, and topography channels are largely decoupled from each other suggesting that the frequency contrast be related to localized regions of higher elastic modulus or stiffness/strain. The size and area coverage of shows similarities with regions of differing crystallinity found in the topography maps (see Figure S2). Importantly the variation in the contact resonance frequency does explain the phase inversion visualized by single frequency PFM (see Figure S1(g)), again highlighting the unsuitability of single frequency PFM. While localized phase contrast may be indicative of ferroelectric, we see no evidence of phase inversion in these regions (see Figure S3), indicating the presence of single polarization state and absence of ferroelectric domain contrast in this region.

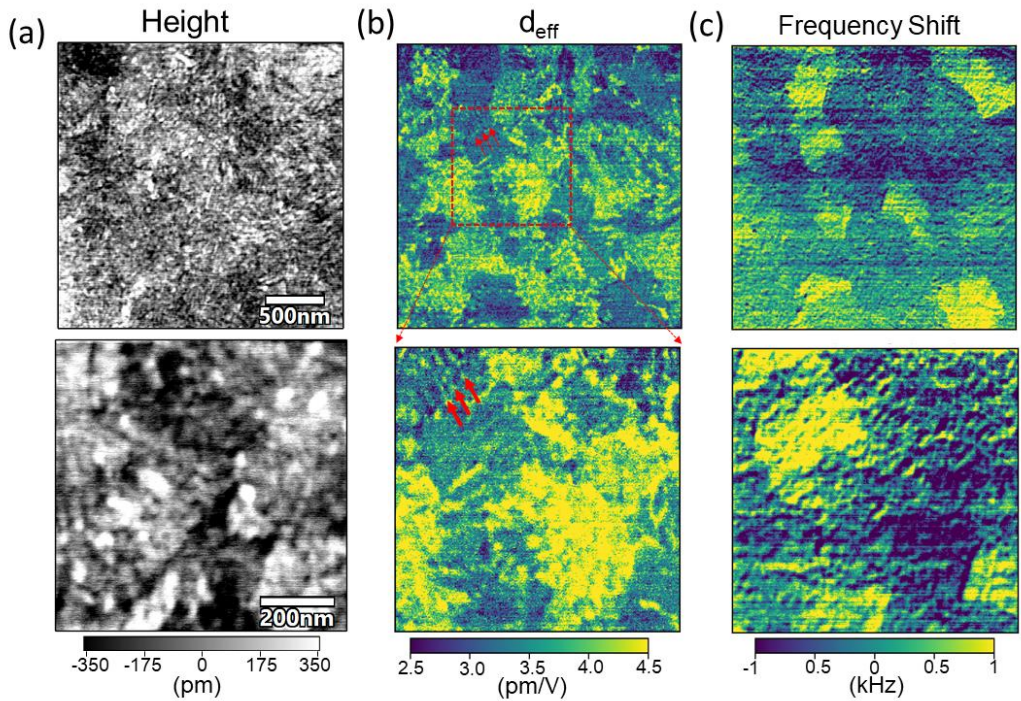

**Figure 2. High resolution BE-PFM of bare HfO$_2$ oxide surface.** (a) Topography scan collected over a 2.5 µm$^2$ region (top) and a smaller 800 nm$^2$ region (bottom). BE-PFM SHO fit

parameters showing (b) the effective piezoelectric coupling coefficient (pm/V) and (c) contact resonance frequency shift. Zoom area is indicated by dashed red box in (b).

In terms of magnitude of the piezoelectricity, averaging across the entire surface gave a $d_{eff}$ value of 3.8 ± 0.7 pm/V. In this case the average value on the bear films are comparable to values measured by PFM on Zr doped $HfO_2$ capacitors having top electrodes (5.5 pm/V).[12] Meanwhile measurements on Y-doped $HfO_2$ measured using double beam interferometry indicated a much lower $d_{eff}$ of 1 pm/V.[25] Unfortunately, while BE-PFM avoids indirect crosstalk with topography information, it is still sensitive to non-local electrostatic interactions largely inhibiting such quantitative comparisons.[26] Indeed, all PFM methods that utilize an optical beam detection approach,[18,19] will be influenced by vibrations of the cantilever arising from non-piezoelectric contributions. Such effects are indistinguishable from local piezoelectric tip displacements and hence detrimental to quantitative measurements. We note that there have been extensive efforts to minimize the background artifacts in OBD based PFM methods, including using high loads,[9] stiff tips,[27,28] long shanks,[27] compensating voltages,[29] as well as mode shape corrections using analytical[30] and experimental methods,[31] with varying degrees of success. Here, to unambiguously quantify the nanoscale piezoelectric properties in silicon-doped $HfO_2$ we adopt a newly developed detection scheme utilizing interferometric detection sensing (IDS) as opposed to OBD.

IDS-PFM utilizes interferometric detection to directly measure the vertical tip displacement, while avoiding parasitic contributions from the electrostatic excitation of the cantilever beam.[32] Further, since IDS is an intrinsically calibrated approach it avoids complications associated with conversion of the OBD slope measurement to a physical measure of tip displacement.[20] For all IDS-PFM measurements the detection spot has been optimized for detection of tip displacements and avoidance of unwanted cantilever motion as described elsewhere.[20, 32] In general both BE-PFM and IDS-PFM showed comparable nanoscale contrast, confirming the nanoscale variability of the piezoelectric properties of bare $HfO_2$ (see comparison images in Figure S4). In contrast between the techniques, we found that BE-PFM consistently measured a higher effective $d_{eff}$ compared to IDS-PFM, and whereas BE-PFM measured a finite $d_{eff}$ everywhere on the film, IDS-PFM measured regions showing no detectable piezoelectric behavior. Indeed, due to the fact the orthorhombic phase is expected to account for ca. 44% of the polycrystalline film, as visible with by X-ray diffraction (not shown), we expect unresponsive regions of the sample. As such we attribute the larger $d_{eff}$ to an artificial shift in background of the BE-PFM signal arising from vibrations of the cantilever due to non-local electrostatic interactions.

Across many measurements, the $d_{eff}$ measured on the bare oxide was consistently found to be reduced with respect to the BE-PFM measurements, giving values between 0.6-1.5 pm/V. It is worth noting that compared to traditional ferroelectric materials (e.g. PZT) the measured coupling is remarkable small, but seemingly in agreement with double beam interferometry measurements.[25] From the IDS-PFM amplitude and phase images in Figure 3, it was possible to clearly distinguish isolated domains of enhanced local electromechanical coupling from inactive regions (or regions with response below the detection noise limit, ~0.4 pm/V). Initially the responsive areas accounted for ~ 26% of the total area covered and apparent 180-degree phase contrast could be distinguished from (Figure 3(f)). The phase channel is noisy due to the small amplitude signal response; however, such phase contrast is often used as an indication of ferroelectric domains. Interestingly, after repeated scanning we did observe a slight increase in the $d_{eff}$ (Figure 3b). From threshold analysis of 3 consecutive scans the responsive areas were found to increase to ~79% of the total area (see Figure S5). The increase in $d_{eff}$ was accompanied by a more uniform phase distribution (Figure 3f-h). Indeed, it is known that increased remnant polarization and loop opening is observed on FE-HfO$_2$ capacitors.[33, 34] Several groups have reported that the underlying cause for this effect might be a redistribution of oxygen vacancies,[34, 35] or a phase transition from a non-FE to a FE phase.[36] Although in this work we have no physical top-electrode, here replaced by our probe, we consider our observation the equivalent of what is commonly observed in fully-fabricated capacitors.

Interestingly, we see clear evidence of localized regions which remain void of piezoelectric response. The inactive regions accounted for ~20% of the total area. Finally, after repeated scanning over the same area, it was possible to identify the previously scanned regions from a larger zoom area. (Figure S6). Similar control over the $d_{eff}$ could be found under the application of dc voltage (Figure S7), however, phase reversal as an indicator of ferroelectricity was not observed.

The anomalously low electromechanical coupling value may be a significant underlying factor behind the lack of convincing reports of ferroelectricity on the nanoscale using PFM on bare film. Unfortunately, the weak electrochemical coupling exhibited by the bare HfO$_2$ surface required the application of moderately large ac voltages (e.g. 2-4V$_{ac}$) to achieve good signal-to-noise values. However, as shown in Figure 1c, no significant leakage is observed in this voltage range for our film. We observed that PFM at these voltages resulted in localized surface charging as revealed by Kelvin probe force microscopy (Figure S8). At larger voltages (i.e. 6 V$_{ac}$), we begin to observe structural changes in the surface oxide as measured by a localized thickness change on the order of ~100 pm (Supplementary Figure S8-9). These observations suggest that surface charging at low bias and non-linear dielectric processes or surface electrochemical effects may impact the $d_{eff}$ measured by PFM, and possible mask underlying ferroic properties.

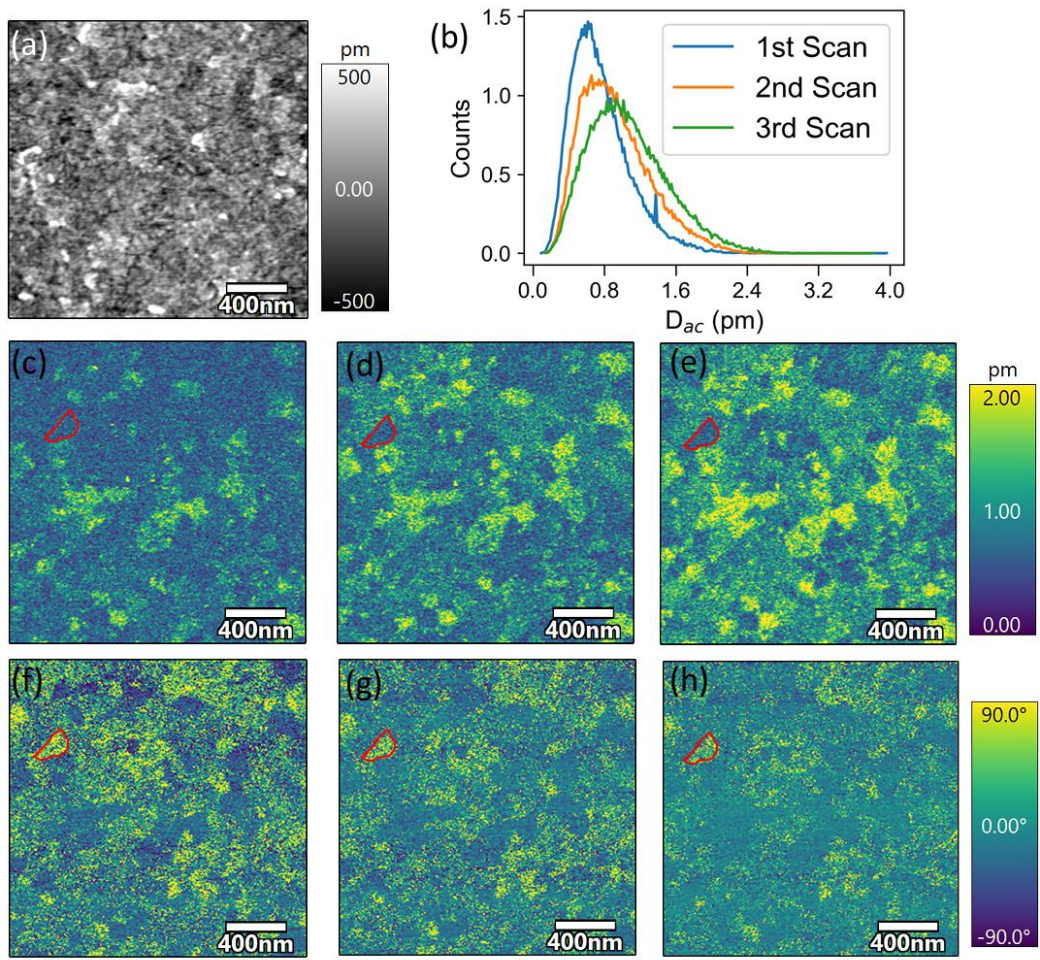

**Figure 3. Repeated IDS-PFM imaging on bare HfO$_2$ film.** (a) Topography scan collected over a 2 µm$^2$ area. (b) Histogram of the measured out of plane piezoresponse over the entire region. Spatial maps of the PFM (c,d,e) amplitude and (f,g,h) response for the (c,f) first, (d,g) second and (e,h) third consecutive scans are shown. Images were collected using a drive amplitude of 4 V$_{ac}$ and drive frequency of 112 kHz.

Finally, in view of the superior immunity of IDS-PFM to electrostatic affects,[20, 32] we performed switching spectroscopy studies on the bare oxide film as in Figure 4. We observed a hysteretic amplitude dependence and a 180-degree phase inversion which is often indicative of ferroelectric switching. Considering the confined volume probed by the tip when used in spectroscopic mode i.e. 500 - 1000 nm$^3$, the results of Figure 4 suggests strong potential for scalability for FE-HfO$_2$-based devices. However, we do note some inconsistencies in the ferroelectric loop shape with respect to a classical ferroelectric material. In particular, we do not observe a saturation of the amplitude loop within the voltage range, which is typical for a classical ferroelectric material. At larger applied voltages irreversible processes were initiated and accompanied by catastrophic topographical breakdown (not shown). Interestingly, we also note a clear and gradual increase in the signal for both biased and remnant measurement as a function of cycling. This is accompanied by a significant broadening of the switching voltage with increased cycling number. The latter is shown in Figure 2(a,b) for one thousand consecutive cycles acquired with a pulse duration of ~10 milliseconds applied between the tip and the sample. The value of the increase in $d_{33}$ observed here, i.e. ~ 30 % in 1000 cycle, which is consistent with the increase in the remnant polarization that is obtained in capacitors based on the same material, as reported elsewhere.[23] While similar loops have been used to directly infer ferroelectricity,[37] the peculiar loop shape might suggest that additional underlying factors (e.g. chemical, electrochemical, and physical phenomena) are involved in the PFM switching of the exposed HfO$_2$ surface.[37] Indeed, similar "ferroelectric-like" loops can be found on non-ferroelectric materials, which have been related to surface charging due to the injection from the tip, charged defects (i.e., oxygen vacancies) migration, electrochemical reactions on the film surface, or other effects.[37] In the case of IDS-PFM we can ignore electrostatic contributions to the signal from surface charging. We further confirmed the absence of non-contact hysteresis (not shown) strongly suggested that the observed IDS measurements are true localized electromechanical response.[32]

Hence we can be confident that the underlying hysteresis is a true localized electromechanical coupling.[32] While difficult to infer from IDS-PFM spectroscopy alone, we suspect that additional non-ferroelectric contributions to the electromechanical coupling arising through either localized ion transport (e.g. oxygen vacancies) under the tip, or interfacial effects,[14] producing a surface displacement which is electrochemical in origin. Interestingly, Cheema et al.[15] also noted similar non-ferroelectric contributions to the switching spectroscopy loops measured by PFM (and IDS-PFM). In this case they speculated that this behavior is related to interfacial effects at the dielectric $SiO_2$ layer through which most of the voltage is dropped. Glinchuk et al., described an electrochemical mechanism for the unusual ferroelectric behavior reported in $HfO_2$ in which they considered oxygen vacancies as elastic dipoles which coupled with the polarization via electrostriction and Vegard effects.[35] Indeed, such mechanisms might be expected from the observations of localized KPFM potentials and oxide growth under ac voltage alone (see SI information).

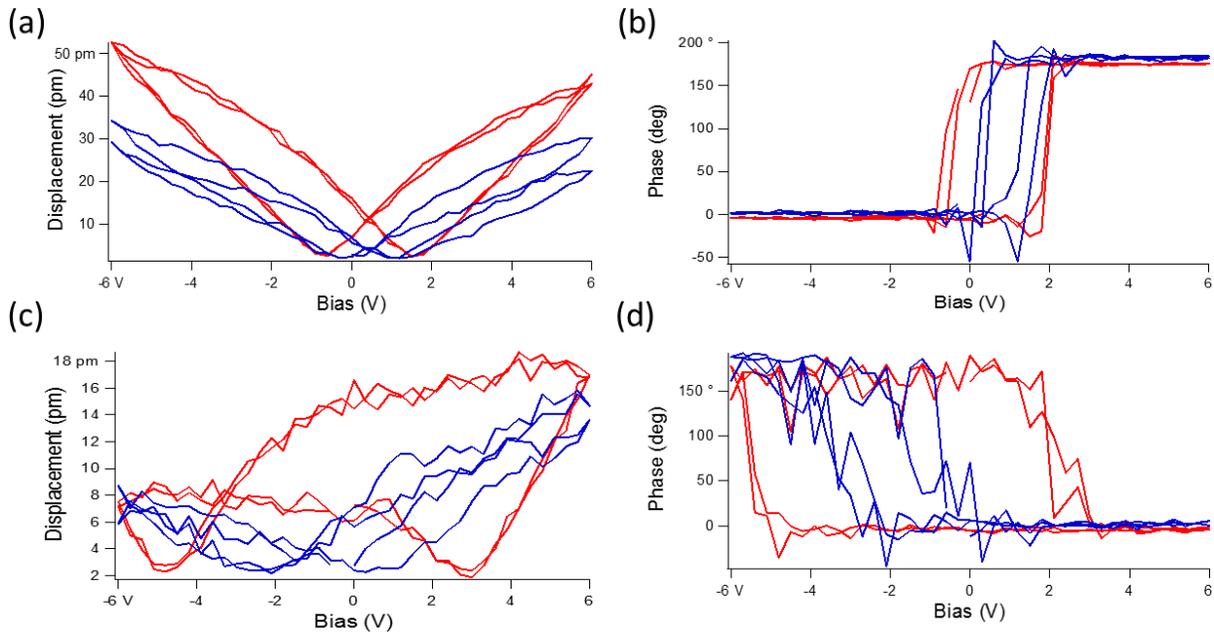

**Figure 4. Switching spectroscopy using IDS-PFM.** Showing (a,c) amplitude and (b,c) phase loops during the (a,b) "bias on" and (c,d) remnant or "bias off" state. Blue curves represent the first switching event whereas the red curve indicates the 1000$^{th}$ switching measurement on the same location.

In summary, local sensing of the effective piezoelectric coefficient on the bare surface of polycrystalline Si-doped HfO$_2$ has been demonstrated. We compare different scanning probe methods including PFM, BE-PFM and IDS-PFM. Thanks to its high sensitivity and immunity to electromechanical sensing artefacts, we conclude that the IDS method offers a quantitative source of information in thin ferroelectric binary oxides. This is especially true as it allows to remove a dc pre-polarization step of the oxide surface, that we prove to induce (Figure S8) a strong source of electrostatic artifacts for PFM and BE-PFM. While our layer exhibits a polycrystalline structure containing multiple grains of different phases, only a small fraction of crystalline grains show contrast using IDS. For these regions we extract an electromechanical coupling coefficient in range 0.6-1.5 pm/V. We suggest that these regions are where the local microstrain in the film is sufficient to stabilize the orthorhombic HfO$_2$ grains. Upon repeated scanning under ac voltage, these locations present also a gradual increase in the sensed converse piezoelectric effect. Consistently, when individual switching spectroscopy is performed on these grains, we observe a visible increase in the amplitude of the loops as a function of repeated cycling. Similar to the interpretation given for macroscopic capacitors, we suggest that the material undergoes a field-induced material activation i.e. phase transition (or domains de-pinning), interestingly in our case even without the presence of interfacial layers and stressors (electrodes) on the top interface. Finally, the nanoscopic domains analysis resulted in a lateral resolution down to 10 nm, thus making IDS and interesting approach for the non-invasive, high resolution quantitative mapping and screening of ferroelectric doped-$HfO_2$ layers.

EXPERIMENTAL METHODS

**AFM measurements**

The AFM used in this study combines a commercial Cypher AFM (Asylum Research, Santa Barbara, CA) with an integrated quantitative Laser Doppler Vibrometer (LDV) system (Polytec GmbH, Waldbronn, Germany) to achieve highly sensitive electromechanical imaging and spectroscopy. All PFM measurements were captured using Pt-coated cantilevers with a spring constant of ∼2 N/m and resonance frequency of ∼75 kHz was used. The Cypher AFM was equipped with external data acquisition electronics based on an NI-6115 fast DAQ card to generate the probing signal and store local hysteresis loops and correlate them with surface topography. BE was achieved by coupling the AFM with external data acquisition electronics based on an NI-6115 fast DAQ card to generate the probing signal and store local BE and hysteresis loops and correlate them with surface topography.

ASSOCIATED CONTENT

**Supporting Information**

Complications associated with single frequency PFM. (Figure S1) Threshold analysis of AFM height measurements on bare Si doped $HfO_2$. Results of the SHO fitting off the contact resonance in BE- PFM (Figure S3). Hysteresis loops captured as a function of distance from a soda-lime glass surface. (Figure S2) DART-VM-AFM measurements on polystyrene/polycaprolactone (PS/PCL) blend. (Figure S5) Influence of repeated PFM scanning on the measured piezoelectric coupling. Larger IDS-PFM scan revealing (Figure S6) AC and (Figure S7) DC induced changes in the localized piezoelectric response. Figure S8 Kelvin probe force microscopy of region where PFM was performed. Figure S9 topography changes after poling with large voltages. (Figure 10) Cross sectional analysis of oxide growth after PFM imaging. This material is available free of charge via the Internet at http://pubs.acs.org.

## AUTHOR INFORMATION

**Corresponding Author**

*E-mail:Collinslf@ornl.gov, Umberto.Celano@imec.be

**Author Contributions**

U.C. synthesized and characterized the Si doped HfO2 device. L.C. carried out acquisition of PFM datasets. L.C. and U.C. participated in discussion and interpretation of results. All authors contributed to the writing of the manuscript.

**Competing Interests**

The authors declare no competing interests.

**Notes**

This manuscript has been authored by UT-Battelle, LLC under Contract No. DE-AC05-00OR22725 with the U.S. Department of Energy. The United States Government retains and the publisher, by accepting the article for publication, acknowledges that the United States Government retains a non-exclusive, paid-up, irrevocable, world-wide license to publish or reproduce the published form of this manuscript, or allow others to do so, for United States Government purposes. The Department of Energy will provide public access to these results of federally sponsored research in accordance with the DOE Public Access Plan (http://energy.gov/downloads/doe-public-access-plan).

ACKNOWLEDGMENT

This research was conducted at and supported (LC) by the Center for Nanophase Materials Sciences, which is a DOE Office of Science User Facility. The authors also want to thank Nina Balke and Rama K. Vasudevan for their valuable discussion. We acknowledge and thank Stephen Jesse for the development of the Band Excitation platform used.

Supplementary Information

# Exploring Nanoscale Ferroelectricity in Doped Hafnium Oxide by Interferometric Piezoresponse Force Microscopy

*Liam Collins[†], and Umberto Celano[ς]*

[†]Center for Nanophase Materials Sciences, Oak Ridge National Laboratory, Oak Ridge, Tennessee 37831, USA

[ς]*IMEC, Kapeldreef 75, B-3001 Heverlee (Leuven), Belgium*

**Frequency dependence in single frequency PFM measurements**

Figure S1(a) shows the topography over which measurements were performed having a roughness of ~205 pm. Figure S1(b) shows a frequency response curve captured at a single location (indicated in (a) with a start) using an AC voltage of 3 V. The data in this case is represented in pm/V where the measured displacement was divided by the driving voltage, which leads to the effective electromechanically coupling coefficient $d_{eff.}$ The frequency dependence for PFM signal collected by optical beam deflection (OBD) is complicated by the contact resonance of the cantilever (~335 kHz) as the OBD method is an indirect measure of the tip displacement via detection of the cantilever motion. It was found that when operated far below resonance using single frequency PFM we measure a mean response of 3.6 ± 1.3 pm/V with no spatial variation in amplitude of phase channels across the sample surface (Figure S1(c,f)). Meanwhile, when operated at the contact resonance frequency it was possible to measure nanoscale contrast ( Figure S1(d,g). In this case the measured "tip" displacements are an order of magnitude larger (234 ± 64 pm/V) due to amplitude by the cantilever flexural mode on resonance (234 ± 64 pm/V). This is not a surprise as we have made no attempt to correct for the amplification due to the contact resonance quality factor. When operated close to the contact resonance we do measure nanoscale contrast, however, the contrast appears dependent on proximity to the contact resonance, where different contrast can be observed when above ( Figure S1(d,g) of below ( Figure S1(e,h) the contact resonance peak.

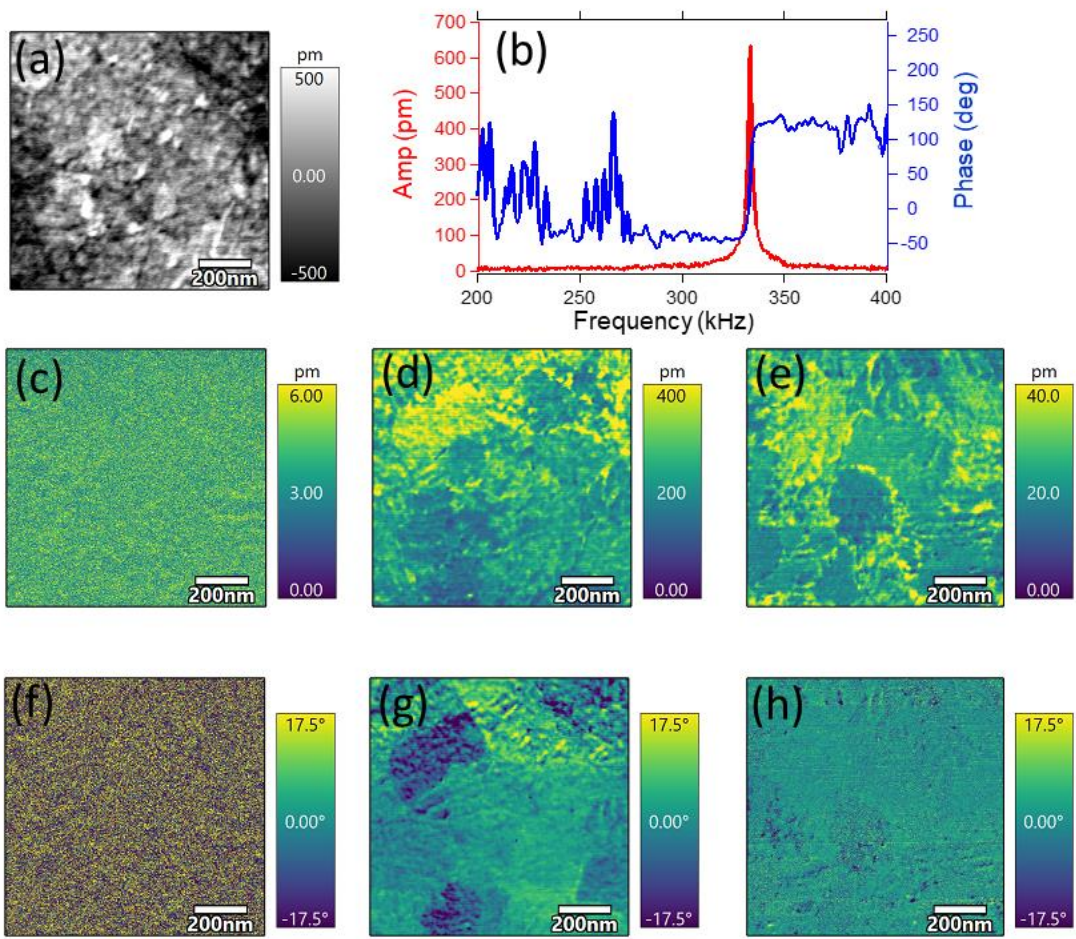

**Figure S1. Single Frequency PFM on bare HfO₂ film.** (a) Topography and (b) single location PFM tune collected by sweeping the frequency of the drive voltage as the tip is held stationary in contact with the oxide film. (c-h) Frequency dependence of the PFM (c-e) amplitude and (f-h) phase captured by single frequency PFM with drive frequencies of (c,f) 215 kHz (below resonance), (d,h) 325 kHz (close to resonance) and (e,h) 350 kHz (above resonance). 3 $V_{ac}$ was used for all measurements and the amplitudes are reported in pm/V by diving the measured amplitude by 3.

**Structural analysis of the bare oxide film**

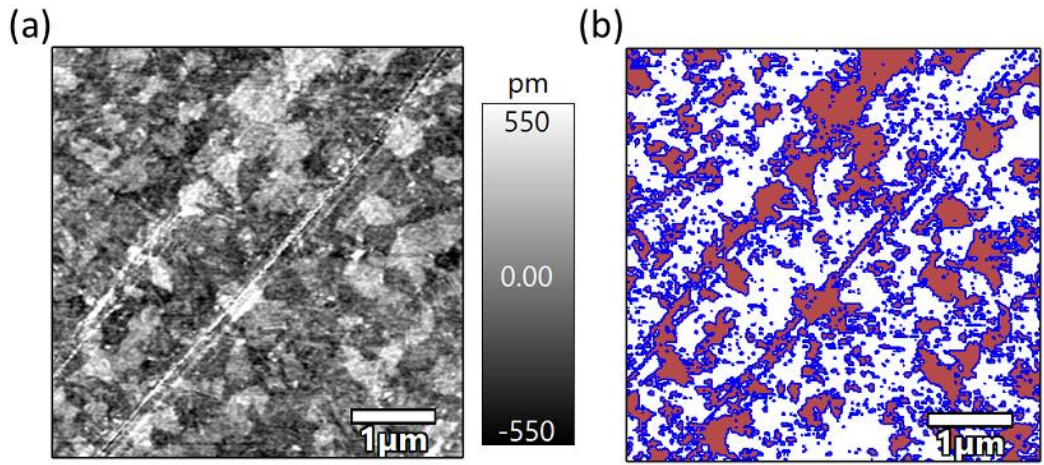

**Figure S2. Threshold Analysis of oxide structure as measured by tapping mode AFM.** (a) Topography scan collected over a 5 x 5 µm area. (b) Threshold mask (red) automatically calculated using an iterative method in igor pro. Higher regions accounted for ~39% of the surface area and were found to be raised by ~200-400 pm with respect to the underlying oxide regions

**SHO fit parameters from BE-PFM**

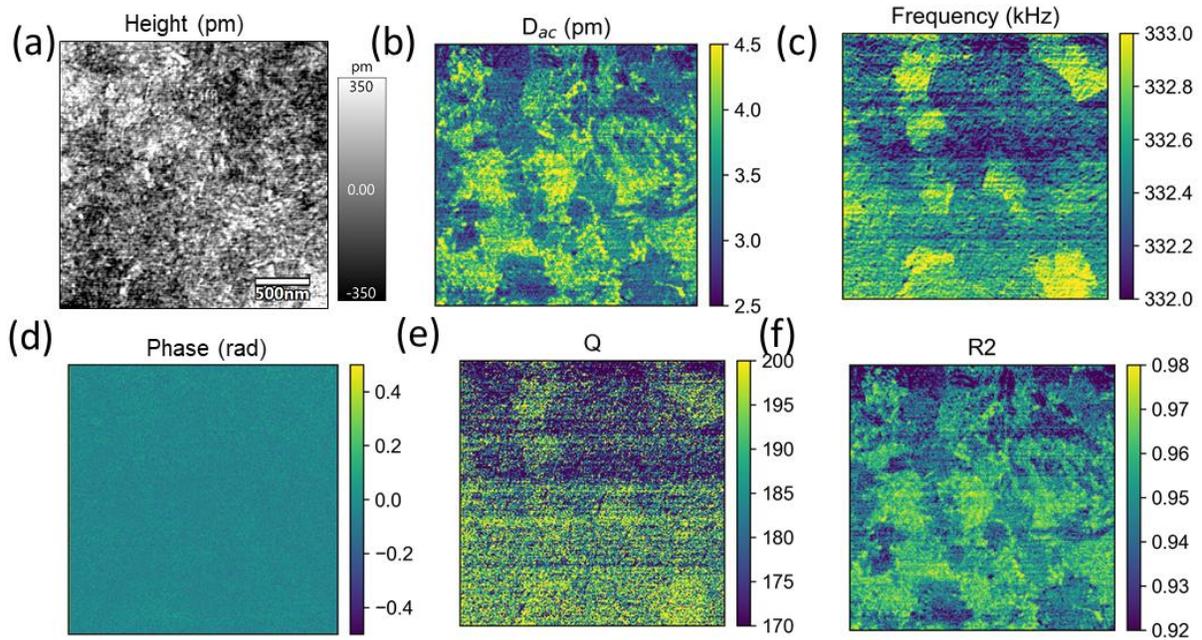

**Figure S3. Band Excitation PFM**. (a) Topography scan collected over a 2.5 x 2.5 μm area. Simple harmonic oscillator fits maps showing the (b) maximum amplitude (pm), (c) contact resonance frequency, (d) phase, (e) quality factor and (f) $R^2$ criteria for the SHO fitting. In contrast to the amplitude and frequency channels, we observe a relatively uniform distribution in the measured Q indicative of little variation in the energy dissipated across the sample. The $R^2$ parameter, which gives a measure of the goodness of SHO fitting. Values greater than 0.9 would indicate that the BE-PFM is well approximated by an SHO model, whereas regions of smaller amplitude show slightly lower $R^2$ values.

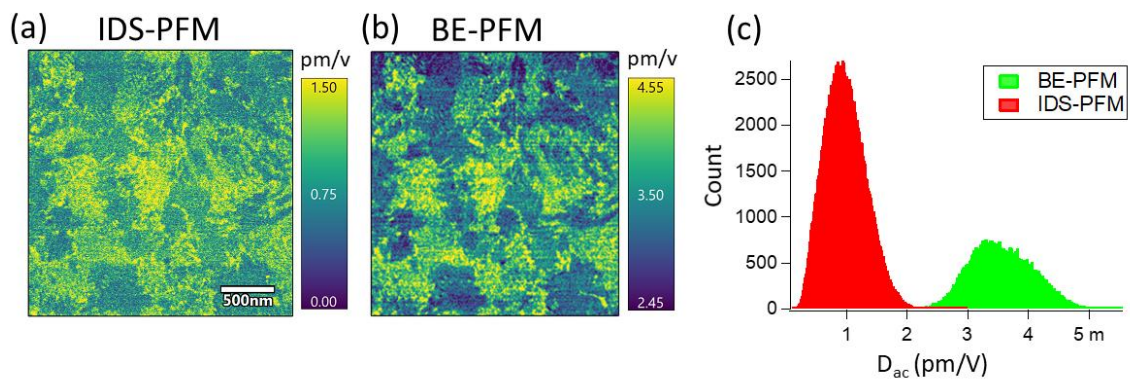

**Figure S4.** Comparison of Piezoresponse measured by (a) IDS-PFM and (b) BE-PFM on the same region of the sample. (c) Histogram of the overall response.

**Influence of applied voltage on measured response**

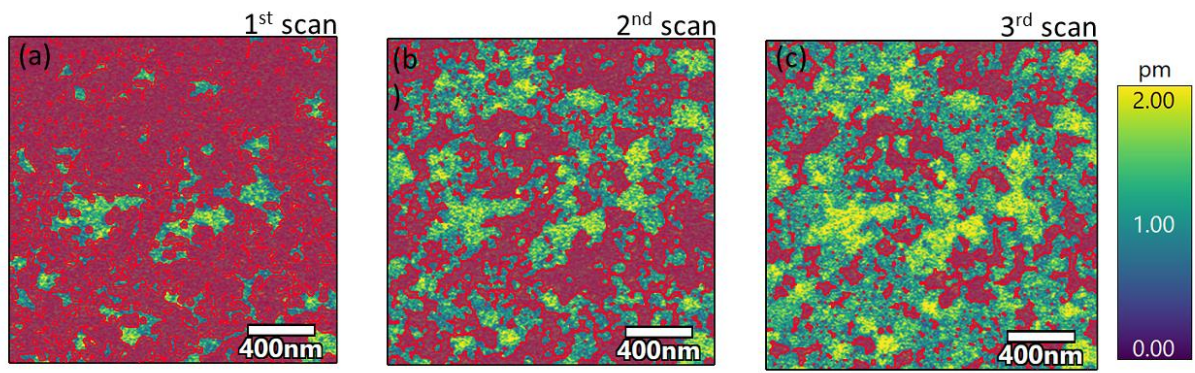

**Figure S5. Threshold analysis of the responsive regions of the sample after consecutive scanning with PFM.** The threshold was set to be the approximate detection noise floor (0.4 pm/V) of the measurement. With repeated scanning the responsive regions accounted for (a) 26% (b) 60 % and (c) 79 % of the surface area.

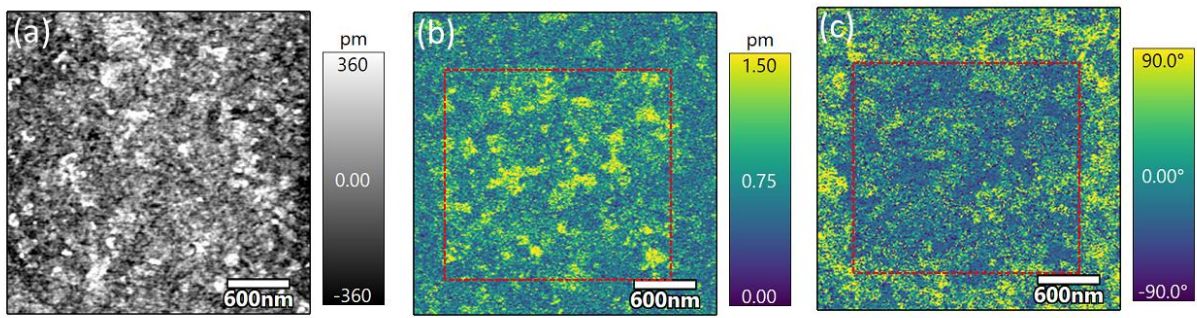

**Figure S6.** Larger IDS-PFM image after repeated scanning with AC voltage applied to the tip. Previously scanned region is shown in (b) and (c). While no evidence of a structural change is observed from the (a) topography image, enhanced (b) piezoreponse and a more uniform (c) phase distribution was found in the previously scanned areas.

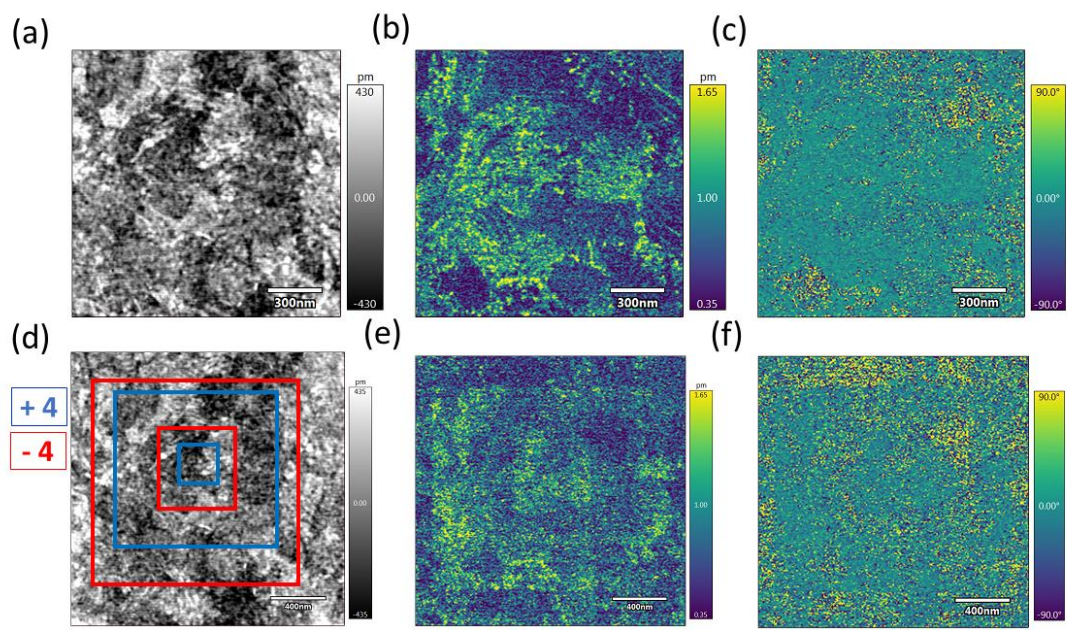

**Figure S7.** Influence of DC voltage applied to the tip. (c,d) Topography, IDS-PFM (d,e) piezoresponse and (c,f) phase images collected (a-c) before and (d,f) after poling experiment involving repeated scanning with ±4V. While no change in topography is observed, clear changes in PFM amplitude and phase is seen in poled regions.

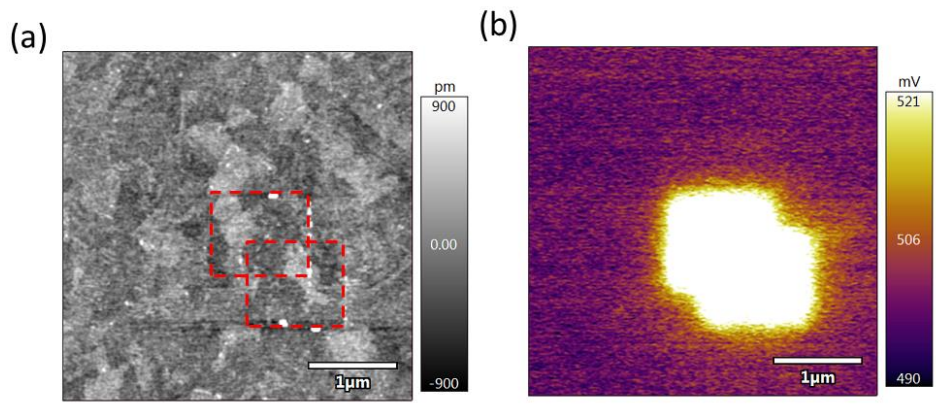

**Figure S8**. KPFM (a) topography and (b) surface potential after scanning region by PFM with an AC voltage of 4V applied to the tip. (red box indicates previously scanned regions shown in Figure 2).

**Structural changes at higher applied bias**

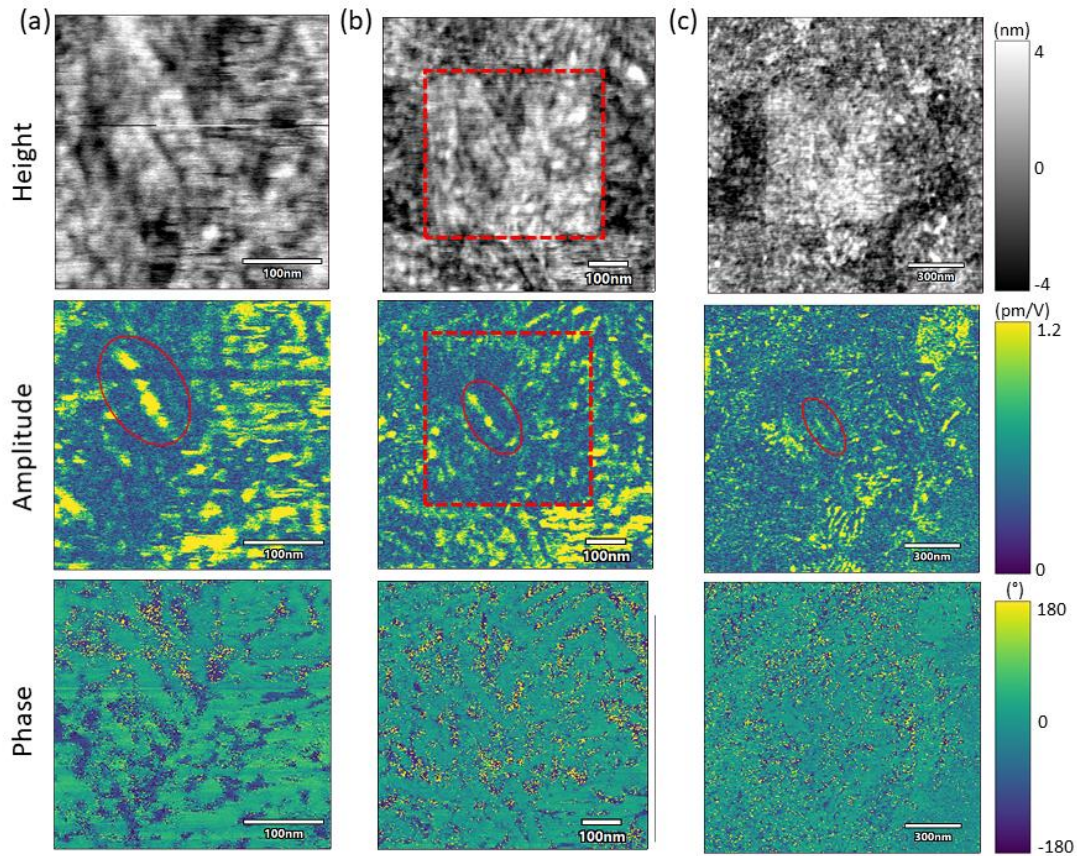

**Figure S9**. IDS-PFM topography (top), Amplitude (middle) and Phase (bottom) for 3 consecutive scans of size (a) 350 nm$^2$, (b) 750 nm$^2$ and (c) 1.5 µm$^2$. Data was collected using a drive frequency of 6V$_{ac}$ at 155 kHz. Scanning was performed at 1Hz line rate and imaging force of ~150 nN.

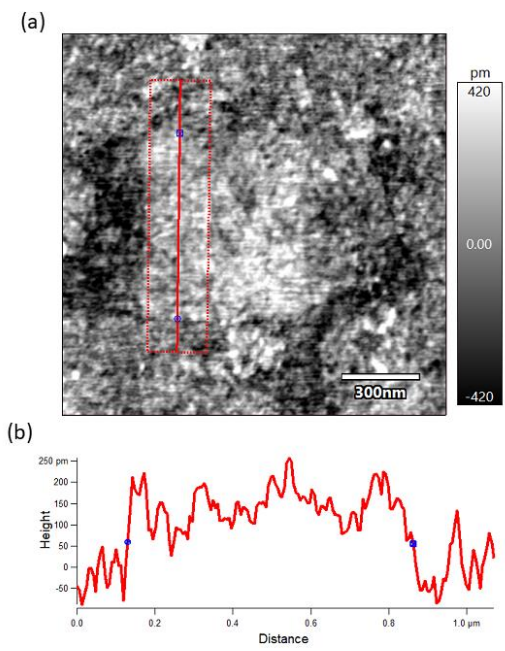

**Figure S10.** Cross sectional analysis of oxide growth after PFM imaging